\documentclass[aip,pop,reprint,superscriptaddress, floatfix]{revtex4-1}

\usepackage{graphicx}
\newcommand{\be}{\begin{displaymath}}
\newcommand{\bn}{\begin{equation}}

\newcommand{\en}{\end{equation}}
\newcommand{\ee}{\end{displaymath}}
\newcommand{\p}{\partial}

\newcommand{\bfm}[1]{\mbox{\boldmath$#1$}}

\usepackage{times}
\usepackage{amssymb}
\usepackage{amsmath}
\usepackage{url,hyperref}

\begin{document}
\title{Collisionless microinstabilities in stellarators I - analytical theory of trapped-particle modes}
\author{P.~Helander}
\author{J.~H.~E.~Proll} 
\affiliation{Max-Planck-Institut f\"ur Plasmaphysik, EURATOM Association,
Teilinstitut Greifswald, Wendelsteinstra{\ss}e 1, 17491 Greifswald, Germany} 
\author{G.~G.~Plunk} 
\affiliation{Max-Planck-Institut f\"ur Plasmaphysik, EURATOM Association,
Teilinstitut Greifswald, Wendelsteinstra{\ss}e 1, 17491 Greifswald, Germany} 
\affiliation{Max-Planck/Princeton Research Center for Plasma Physics} 

\begin{abstract}
This is the first of two papers about collisionless, electrostatic micro-instabilities in stellarators, with an emphasis on trapped-particle modes. It is found that, in so-called {\em maximum-J} configurations, trapped-particle instabilities are absent in large regions of parameter space. Quasi-isodynamic stellarators have this property (approximately), and the theory predicts that trapped electrons are stabilizing to all eigenmodes with frequencies below the electron bounce frequency. The physical reason is that the bounce-averaged curvature is favorable for all orbits, and that trapped electrons precess in the direction opposite to that in which drift waves propagate, thus precluding wave-particle resonance. These considerations only depend on the electrostatic energy balance, and are independent of all geometric properties of the magnetic field other than the maximum-$J$ condition. However, if the aspect ratio is large and the instability phase velocity differs greatly from the electron and ion thermal speeds, it is possible to derive a variational form for the frequency showing that stability prevails in a yet larger part of parameter space than what follows from the energy argument. Collisionless trapped-electron modes should therefore be more stable in quasi-isodynamic stellarators than in tokamaks. 
\end{abstract}

\maketitle
\normalsize
\section{Introduction}
Trapped-electron modes (TEMs) are believed to cause much of the transport observed in tokamaks. These instabilities were predicted theoretically already in the late 1960's \cite{Kadomtsev-1967, Rosenbluth, Kadomtsev-DTEM}, the analytical theory was further developed in the following decades \cite{Kadomtsev-trapped-particles,Adam,Ross,Catto,Rewoldt}, and more recently trapped-particle instabilities have been simulated using gyrokinetic codes. Nearly all the theory and simulations so far have been done for tokamaks, however, with the exception of analytical theory developed for mirror machines in the 1980's \cite{Berk} and a few more recent gyrokinetic simulations for stellarators \cite{Kendl,Pax-2007,Baumgaertel}. Little is therefore known about the properties of TEMs and other trapped-particle instabilities in general magnetic fields, which is the subject of the present work. Our aim is to understand the general characteristics of these instabilities in stellarators, and how serious they are in comparison with their tokamak counterparts. 

Depending on the collision frequency, TEMs are either collisionless or dissipative, and we are mainly concerned with the collisionless branch, which is thought to be particularly harmful in tokamaks. Dissipative TEMs arise whenever there are trapped electrons with high enough collision frequency, but these modes are otherwise fairly independent of the magnetic field, so there is no particular reason to believe that they should be very different in stellarators and tokamaks. The collisionless branch, however, owes its existence to a resonance between drift waves and the precession of trapped electrons \cite{Adam}, and should therefore be sensitive to the geometry of the magnetic field, which determines the direction and magnitude of the trapped-particle precession. Collisionless TEMs could therefore exhibit different behavior in stellarators and tokamaks, and they could also be different in different types of stellarators. As we shall see, this is indeed the case. Anticipating our main result, we note that tokamaks are fundamentally different from stellarators in the sense that the regions of trapping and ``bad curvature'' overlap in tokamaks -- both being on the outboard side of the torus -- whereas in stellarators they may do so, or they may not. Collisionless TEMs are caused by electrons being trapped in regions of bad curvature, and are therfore stabilized in the absence of such an overlap. 

Our work is divided into two parts. The present paper, Part I, is devoted to analytical theory and contains general derviations, valid in arbitrary toroidal magnetic fields, as well as results obtained in analytically tractable limits concerning the magnetic-field geometry. These are approximately, but never exactly, valid in certain stellarators, and it is therefore necessary to confirm the results numerically. This is the task of the second paper, Part II, where the analytical predictions are examined numerically, using a gyrokinetic code. Many of the key analytical results of Part I have recently been reported in condensed form in two previous papers \cite{Proll,Helander}. It is the aim of the present paper to present a more complete picture, providing full mathematical details, and to extend the previous calculations to enable a more accurate comparison with the numerical simulations in Part II. 

\section{Gyrokinetic system of equations}

We consider an arbitrary stellarator with nested magnetic flux surfaces, so that the magnetic field can be written as ${\bf B} = \nabla \psi \times \nabla \alpha$, with $\psi$ the toroidal magnetic flux and $\alpha = \theta - \iota \varphi$ the Clebsch angle, constructed as usual from the poloidal and toroidal magnetic coordinates. In a gyroradius expansion, it follows from the zeroth-order drift kinetic equation that the equilibrium distribution function, $f_{a0}$, of each species $a$ is a Maxwellian at rest \cite{Helander-2007,Sugama-2011}, whose density and temperature are constant on each flux surface. The linear stability of this equilbrium against drift-wave-ordered instabilities is governed by the gyrokinetic equation
	\begin{equation}
	i v_{\|}\nabla_{\|}g_a+(\omega -\omega_{da})g_a = \frac{e_a\phi}{T_a} J_0 \left(\frac{k_\perp v_\perp}{\Omega_a} \right)  \left( \omega - \omega_{*a}^T 
	\right)f_{a0}, 
	\label{gk}
	\end{equation}
in the collisionless and electrostatic approximation. Here $\phi$ is the electrostatic potential perturbation, $J_0$ is a zeroth-order Bessel function, $\Omega_a = e_a B / m_a$ the gyrofrequency, and 
	$$g_a({\bf R},v,\lambda,t)=f_{a1}({\bf r},{\bf v},t) + \frac{e_a\phi({\bf r},t)}{T_a}f_{a0}(v)$$ 
denotes the non-adiabatic part of the perturbed distribution function, which in lowest order becomes independent of the gyroangle when written as a function of the guiding-center position ${\bf R} = {\bf r} - {\bf b} \times {\bf v} / \Omega_a$ rather than the particle position $\bf r$ \cite{Catto}. The parallel derivative is taken at constant magnetic moment $\mu = m_a v_\perp^2 / 2B$, and we shall use $v$ and $\lambda = v_\perp^2 / (v^2 B)$ as our independent velocity-space variables. In addition to the mode frequency $\omega$, two characteristic frequencies appear in Eq.~(\ref{gk}), the drift frequency $\omega_{da} = {\bf k} \cdot {\bf v}_{da}$, and the diamagnetic frequency $\omega_{*a} = (T_a k_{\alpha}/e_a) d \ln n_a / d \psi$ appearing in 
	$$ \omega_{*a}^T = \omega_{*a} \left[ 1 + \eta_a \left(x^2 - \frac{3}{2} \right) \right], $$
with $x^2 = m_a v^2 / 2 T_a$. Here, $n_a$ denotes the density, $T_a$ the temperature, $\eta_a = d \ln T_a / d \ln n_a$, and the wave vector has been written as ${\bf k}_\perp = k_\psi \nabla \psi + k_\alpha \nabla \alpha$. 
The system of equations is closed by the quasineutrality condition, 
\begin{equation}
\sum_a \frac{n_a e_a^2}{T_a}\phi = \sum_a e_a \int g_a J_0 \mathrm{d}^3 v.
\label{qn}
\end{equation}

Although this system of equations is standard, two comments are in order. First, we have taken all perturbuations to be proportional to $e^{iS}$, where the eikonal $S$ does not vary along the magnetic field, so that the wave vector ${\bf k}_\perp = \nabla S$ is perpendicular to $\bf B$. In toroidal configurations, this is possible if the magnetic field lines close on themselves, or if the ballooning transformation is used. 
The latter was adapted to stellarators by Dewar and Glasser \cite{Dewar-Glasser}, and a short summary can be found in Ref.~\cite{Helander}, but it cannot be used when the (global) magnetic shear is very small. The second remark concerns the equilibrium electric field, ${\bf E}_0 = - \nabla \phi_0(\psi)$. In a tokamak, such a field (but not its shear) can be eliminated by transforming to a toroidally rotating frame \cite{Helander-Sigmar} and therefore plays no role for microinstabilities in the gyrokinetic ordering, but it is perhaps not obvious that it should be unimportant in a stellarator. That this is the case follows, however, quickly from the original formulation of the gyrokinetic equation \cite{Antonsen-Lane,Catto-Tang-Baldwin} retaining an equilibrium electric field of order $e \phi_0 / T = O(1)$, which simply has the effect of Doppler-shifting the frequency $\omega$. 

Apart from the usual assumptions in gyrokinetics, two approximations have been made in Eq.~(\ref{gk}): electromagnetic effects and collisions have been neglected. The former are unimportant in the limit $\beta \rightarrow 0$, but it is in practice difficult to know {\em a priori} just how small $\beta$ needs to be (typically below one or a few percent, depending on the magnetic geometry).  Collisions are negligible as long as the collision frequency is smaller than $\omega / f_t^2$, where $f_t$ denotes the fraction of trapped particles \cite{Connor-TEM}.

\section{Conventional tokamak approximation}

The conventional way of analytically calculating collisionless TEM stability in tokamaks is to make two basic approximations. The parallel phase velocity of the instability is taken to be intermediate between the ion and electron thermal speeds,
	\bn k_\| v_{Ti} \ll \omega \ll k_\| v_{Te}, 
	\label{drift-wave ordering}
	\en
so as to avoid strong Landau damping on either species, and the inverse aspect ratio is assumed to be small, $\epsilon \ll 1$, implying that the magnetic drift frequency is much smaller than the diamagnetic frequency, 
	\bn \frac{\omega_{da}}{\omega_{*a}} \sim \epsilon \ll 1, \label{small drift}
	\en
for all species. Here we have taken the density gradient length scale to be of the order of the minor radius and the radius of curvature equal to the major radius. In addition, the fraction of trapped particles then becomes small,
	\bn f_t \sim \sqrt{ \epsilon} \ll 1. \label{ft}
	\en
The ordering (\ref{drift-wave ordering}) makes it possible to solve the gyrokinetic equation (\ref{gk}) very easily. For the ions, the first term on the left can simply ignored, giving
	\bn g_i = \frac{\omega - \omega_{*i}^T}{\omega - \omega_{di}} \frac{e J_0 \phi}{T_i} \; f_{io}, 
	\label{gi}
	\en
which reduces to 
	\bn g_i \simeq \left( 1 - \frac{\omega_{*i}^T}{\omega} \right) \frac{e J_0 \phi}{T_i} \; f_{io}, 
	\label{gi0}
	\en
because of the approximation (\ref{small drift}), since the frequency $\omega$ will turn out to be of order $\omega_{*i}$. For the electrons, the distribution function is expanded, $g_e = g_{e0} + g_{e1} + \cdots$, giving $g_{e0} = 0$ in the circulating part of velocity space and
	\bn g_{e0}^{\rm tr} = - \frac{\omega - \omega_{*e}^T}{\omega - \overline{\omega}_{de}} \frac{e \overline{\phi}}{T_e} \; f_{eo} 
	\label{ge}
	\en
in the trapped region. Here, an overbar denotes a bounce average for trapped particles, 
	$$ \overline{\phi}(\lambda) = \int \frac{\phi(l) \; dl}{\sqrt{1 - \lambda B(l)}} \bigg\slash \int \frac{dl}{\sqrt{1 - \lambda B(l)}}, $$
where the integrals are taken along the magnetic field between two consecutive bounce points, defined by $\lambda B = 1$. Since the fraction of trapped particles is assumed to be small, the quasineutrality condition (\ref{qn}) reduces to 
	$$ \left(\frac{1}{T_e} + \frac{1}{T_i}\right) \phi = \frac{1}{ne} \int g_i J_0 \; d^3v $$
in lowest order. Using the approximation (\ref{gi0}) for $g_i$ then gives the drift-wave frequency \cite{Adam,Connor-TEM}
	\bn \frac{\omega}{\omega_{*e}} = \frac{\Gamma_0 - \eta_i b (\Gamma_0 - \Gamma_1)}{1 + \frac{T_e}{T_i} (1 - \Gamma_0)}, 
	\label{omega0}
	\en
where $\Gamma_n(b) = e^{-b} I_n(b)$ and $b = k_\perp^2 T_i / m_i \Omega_i$. In the limits of long and short wavelength (compared with the ion gyroradius), respectively, this frequency is
	\bn \frac{\omega}{\omega_{*e}} \rightarrow 1, \qquad b \rightarrow 0, 
	\label{low b}
	\en
and 
	\bn \frac{\omega}{\omega_{*e}} \rightarrow \frac{1 - \eta_i/2}{\left( 1 + \frac{T_e}{T_i} \right) \sqrt{2 \pi b}}, \qquad b \rightarrow \infty. 
	\label{high b}
	\en
Note that the frequency can be of either sign, in general. For modest ion temperature gradients, $\eta_i < 1.64$, $\omega / \omega_{*e}$ is however always positive (see Fig.~1 below), so that the mode propagates in the electron diamagnetic direction. 

The next-order correction to the dispersion relation becomes
\begin{align}
 \phi \int \left( \frac{\omega + \delta \omega - \omega_{*i}^T}{\omega + \delta \omega - \omega_{di}}
	- \frac{\omega - \omega_{*i}^T}{\omega} \right) J_0^2 f_{i0} \; d^3v
	& \notag \\ 
 +  \frac{T_i}{T_e} \int_{\rm tr.} \frac{\omega - \omega_{*e}^T}{\omega - \overline{\omega}_{de}} \; \overline{\phi} f_{e0} \; d^3v & = 0, 
	\label{correction}
\end{align}
where we have now denoted the zeroth-order frequency by $\omega$ and the first-order correction by $\delta \omega$ (still ignoring effects due to finite $k_\| v_{Ti} / \omega$). The latter acquires an imaginary part from the resonant denominators, leading to the collisionless TEM. The traditional way of estimating the growth rate is to take $\overline{\phi} \simeq \phi$ and ignore the ion resonance \cite{Adam,Connor-TEM}, giving
	$$ \frac{\delta \omega}{\omega^2} \int \omega_{*i}^T J_0^2 f_{i0}\; d^3v
\simeq -  \frac{T_i}{T_e} \int_{\rm tr.} \frac{\omega - \omega_{*e}^T}{\omega - \overline{\omega}_{de}} \;
	f_{e0} \; d^3v $$
and hence, since $\gamma \ll \omega$, 
	$$ \frac{\gamma}{\omega}
	\simeq \frac{\pi / n}{1 + \frac{T_e}{T_i} (1 - \Gamma_0)} \int_{\rm tr.} (\omega - \omega_{*e}^T) \delta(\omega - \overline{\omega}_{de})
	f_{e0} \; d^3v  . $$
If the real frequency $\omega$ has the same sign as $\overline{\omega}_{de}$, this dispersion relation predicts instability with an exponentially small growth rate, $\gamma \sim \exp(- \omega / \overline{\omega}_{de})$. 

There is no rigorous justification for these approximations, and we shall instead assess the stability from more general energy-balance arguments below. Note that both the ions and the electrons have the potential of destabilizing the mode, through their respsective resonant denominators in Eq.~(\ref{correction}).

\section{Large-aspect-ratio stellarator approximation}

\subsection{Integral equation}

Stellarators typically have large aspect ratio, but the number of trapped particles is nevertheless not necessarily small. In W7-X, for instance, the magnetic field spectrum has a strong toroidal mirror component and the fraction of trapped particles is several tens of percent on the magnetic axis. Thus, although the approximations (\ref{drift-wave ordering}) and (\ref{small drift}) may be appropriate, Eq.~(\ref{ft}) is certainly not. We  therefore need to take the trapped-electron response (\ref{ge}) into account already in lowest order, so that the quasineutrality equation (\ref{qn}) becomes
\begin{align} 
\left(1 + \frac{T_e}{T_i} \right) \phi = & \frac{\phi T_e}{n T_i}
	\int \frac{\omega - \omega_{*i}^T}{\omega - \omega_{di}}  J_0^2f_{i0} d^3v\notag \\
	& + \frac{1}{n} 
	\int \frac{\omega - \omega_{*e}^T}{\omega - \overline{\omega}_{de}} H(\lambda) \overline{\phi} f_{e0} d^3v,
	\label{integral eq0}
\end{align}
where $H(\lambda)$ denotes a Heaviside function that is equal to unity in the trapped region, $1/B_{\rm max} < \lambda < 1/B_{\rm min}$ and vanishes in the circulating region, $\lambda < 1/B_{\rm max}$. Here $B_{\rm min}$ and $B_{\rm max}$ denote the smallest and the largest magnetic field strength on the flux surface under consideration. 

At the point(s) along the field line where $B = B_{\rm max}$, there are no trapped particles and the the second integral in Eq.~(\ref{integral eq0}) vanishes. It follows from this equation, then, that there are two possibilities at each such point: $\phi$ either vanishes or $\omega$ satisfies the dispersion relation 
	$$ 1 + \frac{T_i}{T_e} = \frac{1}{n}
	\int \frac{\omega - \omega_{*i}^T}{\omega - \omega_{di}}  J_0^2f_{i0} d^3v $$
The first possibility corresponds to the TEM and the second to the toroidal ion-temperature-gradient mode. In the present paper, we focus on the former.  

\subsection{Zero-magnetic-drift approximation}

Ignoring the magnetic drift frequency in Eq.~(\ref{integral eq0}) by invoking the ordering (\ref{small drift}) gives the dispersion relation
\begin{align}
 & \left[ 1 + \frac{T_e}{T_i}(1 - \Gamma_0) - \frac{\omega_{*e}}{\omega} \left( \Gamma_0 - \eta_i b (\Gamma_0 - \Gamma_1) \right) \right] \phi \notag\\
= & \left( 1 - \frac{\omega_{*e}}{\omega} \right) \frac{B}{2} \int_{1/B_{\rm max}}^{1/B} \frac{\overline{\phi} \; d\lambda}{\sqrt{1- \lambda B}}, 
	\label{integral eq}
\end{align}
in the form of an integral equation for $\phi$, where the left-hand side represents the earlier dispersion relation (\ref{omega0}). It does not seem possible to solve this integral equation analytically, but it is possible to reformulate it as a variational principle, where the variational quantity is equal to the mode frequency $\omega$. This is accomplished by multiplying Eq.~(\ref{integral eq}) by $\phi^\ast / B$ and integrating along the entire field line, using
	$$ \int_{-\infty}^\infty \phi^\ast(l) dl \int_{1/B_{\rm max}}^{1/B} \frac{\overline{\phi}_j \; d\lambda}{\sqrt{1- \lambda B}}
	= \int_{1/B_{\rm max}}^{1/B_{\rm min}} \sum_j \tau_j | \overline{\phi}_j |^2 d\lambda, $$
where the sum is taken over all relevant magnetic wells, i.e., over all regions with magnetic field strength $B < 1/\lambda$, and 
	$$ \overline{\phi}_j(\lambda) = \frac{1}{\tau_j(\lambda)} \int \frac{\phi(l) \; dl}{\sqrt{1 - \lambda B(l)}}, $$
denotes the bounce average of $\phi$ over the j'th such well, with
	$$ \tau_j(\lambda) = \int \frac{dl}{\sqrt{1 - \lambda B(l)}}. $$
It should perhaps be pointed out that, in ballooning space, there is an infinite number of trapping wells along the field line. These are arranged periodically in the tokamak and aperiodically in a stellarator. 
Hence we obtain the following expression for $\omega$, 
	\bn \frac{\omega}{\omega_{*e}} = \frac{N[\phi]}{D[\phi]}, 
	\label{variational form}
	\en
where the functionals $N$ and $D$ are defined by 
\begin{align*}
 N[\phi] = &\int_{-\infty}^\infty \left[\Gamma_0 - \eta_i b (\Gamma_0 - \Gamma_1)\right] |\phi |^2 \frac{dl}{B} \\
	&- \frac{1}{2} \int_{1/B_{\rm max}}^{1/B_{\rm min}} \sum_j \tau_j | \overline{\phi}_j |^2 d\lambda, \\
&\\
D[\phi] = &\int_{-\infty}^\infty \left[ 1 + \frac{T_e}{T_i} (1 - \Gamma_0) \right] |\phi |^2 \frac{dl}{B} \\
	&- \frac{1}{2} \int_{1/B_{\rm max}}^{1/B_{\rm min}} \sum_j \tau_j | \overline{\phi}_j |^2 d\lambda. 
\end{align*}
The denominator $D[\phi]$ is always positive, since the Schwarz inequality, $| \overline{\phi} |^2 \le \overline{| \phi |^2}$, implies
	$$ \frac{1}{2} \int_{1/B_{\rm max}}^{1/B_{\rm min}} \sum_j \tau_j | \overline{\phi}_j |^2 d\lambda \le 
	\frac{1}{2} \int_{1/B_{\rm max}}^{1/B_{\rm min}} \sum_j d\lambda \int \frac{|\phi^2| dl}{\sqrt{1-\lambda B}} $$
	$$ = \frac{1}{2} \int |\phi |^2 dl \frac{1}{2} \int_{1/B_{\rm max}}^{1/B} \frac{d\lambda}{\sqrt{1 - \lambda B}} 
	\le \sqrt{1-\frac{B_{\rm min}}{B_{\rm max}}} \int |\phi |^2 \frac{dl}{B}. $$
	 	
The expression (\ref{variational form}) is variational and assumes its minimum for the particular function $\phi(l)$ that satisfies the integral equation (\ref{integral eq}), as follows from the fact that the variation vanishes,
	$$ \frac{\delta \omega}{\omega} = \frac{\delta N}{N} - \frac{\delta D}{D} = \frac{1}{N} \left( \delta N - \frac{\omega}{\omega_{*e}} \delta D \right) = 0 $$
if, and only if, the integral equation (\ref{integral eq}) is satisfied. Thus, instead of having to solve the integral equation (\ref{integral eq}), we expect that a good approximation to $\omega$ can be obtained by inserting an appropriate trial function in the expression (\ref{variational form}). The traditional systematic way of doing this (Rayleigh-Ritz optimization) is to use a trial function containing one or several free parameters, $\phi(l,\lambda_1, \lambda_2, \ldots)$, and to minimize the variational form with respect to these. This means that one avoids having to solve an integral equation, and in addition one obtains the eigenvalue with enhanced precision: if an error of order $\delta$ is made in the trial function, the ensuing error in $\omega$ is of order $\delta^2$.

\subsection{Sinusoidal wells}

As an example, we consider the simplest case where the magnetic field strength varies sinusoidally along the field line within each trapping well,
	$$ B(l) = B_0 - B_1 \cos (l/L). $$
The simplest possible trial function, without free parameters, which has the required property of vanishing at the field maxima is
	$$ \phi(l) = \frac{\phi_0}{2} \left( 1 + \cos \frac{l}{L} \right). $$
Then
	$$ \tau(\lambda) = \frac{4L}{\sqrt{2 \lambda B_1}} K(m), $$
	$$ \overline{\phi}(\lambda) = \frac{E(m)}{K(m)} \phi_0, $$
where $K$ and $E$ are elliptic integrals of the argument
	$$ m = \frac{1 - \lambda (B_0 - B_1)}{2 \lambda B_1}, $$
and we obtain
	$$ \frac{1}{2} \int_{1/B_{\rm max}}^{1/B_{\rm min}} \tau | \overline{\phi} |^2 d\lambda
	= \frac{2 \phi_0^2 L}{B_0 - B_1} I\left(\frac{2 B_1}{B_0 - B_1} \right), $$
where the function $I$ is defined by
	$$ I(x) = \sqrt{x} \int_0^1 \frac{E^2(m)}{K(m)} \frac{dm}{(1 + mx)^{3/2}} 
	\simeq 0.97 \sqrt x + O\left(x^{3/2}\right). $$
This result will be used in Part II. For the moment, we note that, for shallow magnetic wells, $B_1 \ll B_0$ the correction to the eigenmode frequency (\ref{variational form}) that arises from trapped particles is proportional to the square root of the well depth $B_1/B_0$. Indeed, since to zeroth order in $B_1 / B_0$, 
	$$ \int_{-L}^L \phi^2 \frac{dl}{B} = \frac{3 \pi}{4} \frac{\phi_0^2 L}{B}, $$
the predicted eigenfrequency (\ref{variational form}) becomes
	$$ \frac{\omega}{\omega_{*e}} 
	= \frac{\Gamma_0 - \eta_i b (\Gamma_0 - \Gamma_1) - 1.17 \sqrt{B_1/B_0}}
	{1 + \frac{T_e}{T_i} (1 - \Gamma_0) - 1.17 \sqrt{B_1/B_0}} + O\left( \frac{B_1}{B_0} \right). $$
For what comes later, the most important effect of the finite-trapping correction is that it can make the sign of the numerator change, and thus reverse the direction in which the mode propagates. 

\subsection{Correction due to finite magnetic drift frequency}

The expression (\ref{variational form}) does not contain $\eta_e$ and thus predicts mode frequencies that are independent of the electron temperature gradient. To capture this dependence, it is necessary to account for corrections due finite values of $\overline{\omega}_{de} / \omega  \ll 1$. Still ignoring the resonance, we thus expand
	$$ \frac{\omega - \omega_{*e}^T}{\omega - \overline{\omega}_{de}} 
	= \left(1 - \frac{\omega_{*e}^T}{\omega} \right)\left(1 + \frac{\overline{\omega}_{de}}{\omega} \right) 
	+ O\left(\frac{\overline{\omega}_{de}^2}{\omega^2} \right). $$
Writing $\overline{\omega}_{de} = \tilde \omega_{de}(\lambda) x^2$, we thus obtain the following electron contribution to Eq.~(\ref{integral eq0}),
\begin{align*}
& \frac{2B}{\sqrt{\pi}} \int_0^\infty  e^{-x^2} x^2 dx \int_{1/B_{\rm max}}^{1/B} 
	\frac{\omega - \omega_{*e}^T}{\omega - \overline{\omega}_{de}} 
	\frac{\overline{\phi} \; d\lambda}{\sqrt{1-\lambda B}}  \\
	=& B \int_{1/B_{\rm max}}^{1/B} g(\omega,\lambda)
	 \frac{\overline{\phi} \;  d\lambda}{\sqrt{1 - \lambda B}}, 
	 \end{align*}
where
	$$ g(\omega,\lambda) = \frac{1}{2} \left[1 - \frac{\omega_{*e}}{\omega} 
	+ \frac{3 \tilde \omega_{de}}{2\omega} \left(1 - \frac{(1+\eta_e) \omega_{*e}}{\omega} \right) \right]. $$
Similarly expanding the contribution from $\omega_{di} / \omega \ll 1$ gives for the ion term in Eq.~(\ref{integral eq0}) 
	$$ \frac{\phi T_e}{nT_i}
	\int \frac{\omega - \omega_{*i}^T}{\omega - \omega_{di}}  J_0^2f_{i0} d^3v
	= \frac{\phi T_e}{n T_i} h(\omega,l), $$
with
\begin{align*} h(\omega,l) = & \Gamma_0(b) \left[ 1 - \frac{\omega_{*i}}{\omega} + \frac{\hat \omega_{di}}{\omega}
	- \frac{(1 + \eta_i) \omega_{*i} \hat \omega_{di}}{\omega^2} \right.\\
	& \left.+ b \left( \frac{\eta_i \omega_{*i}}{\omega} - \frac{\hat \omega_{di}}{2 \omega} 
	+ \frac{5 \eta_i \omega_{*i} \hat \omega_{di}}{2 \omega^2} \right) 
	- b^2 \frac{\eta_i \omega_{*i} \omega_{di}}{\omega^2} \right]\\ 
		&+ \Gamma_1(b) \left[ - \frac{\eta_i \omega_{*i}}{\omega}
	+ \frac{b \hat \omega_{di}}{2 \omega} \left( 1 - \frac{\omega_{*i}}{\omega} \right) \right.\\
	& \left.- \left( \frac{3}{2} - b \right) \frac{\eta_i \omega_{*i} \hat \omega_{di}}{\omega^2} \right] 
	\end{align*}
When the effect of a small but finite magnetic drift frequency is taken into account, the integral equation (\ref{integral eq}) is thus replaced by an equation of the form
	\bn f(\omega,l) \phi(l) = B \int_{1/B_{\rm min}}^{1/B} g(\omega,\lambda) \overline{\phi} (\lambda) \frac{ d\lambda}{\sqrt{1-\lambda B}},
	\label{integral eq2}
	\en
with
	$$ f(\omega,l) = 1 + \frac{T_e}{T_i} \left[ 1 - h(\omega, l)\right]. $$

Again, we can reformulate this integral equation as a variational principle, of a somewhat less conventional form than Eq.~(\ref{variational form}). Multiplying Eq.~(\ref{integral eq2}) by $\phi^\ast /B$ and integrating along the entire field line gives a quadratic equation,
\begin{align} S[\phi,\omega] & \equiv
	\int_{-\infty}^\infty f(\omega,l) |\phi|^2 \frac{dl}{B} - \int_{1/B_{\rm max}}^{1/B_{\rm min}} \sum_j \tau_j g(\omega,\lambda) | \overline{\phi}_j |^2 d\lambda\notag\\
	& = 0, 
	\label{S}
\end{align}
for the frequency $\omega$ if the mode structure $\phi$ is known. If the latter is varied, so that $\phi$ is replaced by $\phi + \delta \phi$, then the corresponding change in the frequency, $\delta \omega$, is given by the equation 
\begin{align*}
\delta S[\phi,\omega] = & \int_{-\infty}^\infty \left( \frac{\p f}{\p \omega} \delta \omega |\phi|^2 
	+ 2 f \phi \delta \phi\right) \frac{dl}{B}\\
	 & - \int_{1/B_{\rm max}}^{1/B_{\rm min}} \sum_j \tau_j \left( \frac{\p g}{\p \omega} \delta \omega | \overline{\phi}_j |^2 
	+ 2 g \overline{\phi} \delta \overline{\phi}  \right) d \lambda \\
	 =& 0, 
	\end{align*}
which can be written as 
\begin{equation*}
\delta \omega = \frac{- 2 \int_{-\infty}^\infty \delta \phi \; \frac{dl}{B} 
	\left( f \phi - 
	B \int_{1/B_{\rm max}}^{1/B} \frac{g \overline{\phi}  d\lambda}{\sqrt{1-\lambda B}} \right)}
	{\left( \int_{-\infty}^\infty \frac{\p f}{\p \omega} |\phi|^2 \frac{dl}{B}
	- \int_{1/B_{\rm max}}^{1/B_{\rm min}} \sum_j \tau_j \frac{\p g}{\p \omega} | \overline{\phi}_j |^2 \right)}. 
\end{equation*}
Hence it follows that $\delta \omega = 0$ if the integral equation (\ref{integral eq2}) is satisfied by the pair $(\omega,\phi)$. Conversely, if $\delta \omega = 0$ for all variations $\delta \phi$, then Eq.~(\ref{integral eq2}) is satisfied. The latter is thus, in this sense, equivalent to a variational principle. This variational property can again be utilized within a Rayleigh-Ritz optimization procedure. One substitutes a suitable trial function, $\phi(l,\lambda_1, \lambda_2, \ldots)$, in the definition (\ref{S}) of the functional $S[\phi,\omega ]$, which then becomes a function of $\omega$ and the parameters $\lambda_j$. The system of equations
	$$ S(\omega, \lambda_1, \lambda_2, \cdots) = 0, $$
	$$ \frac{\p S}{\p \lambda_j} = 0, $$
then produces an approximate solution (for both $\phi$ and $\omega$) to the eigenvalue problem (\ref{integral eq2}). As before, higher accuracy is attained for the frequency than for the eigenfunction.

\section{Stability and electrostatic energy balance}

As already mentioned, the modes under consideration can acquire a finite growth rate, $\gamma = {\rm Im} \; \omega$, through the resonant denominators present in the integral equation (\ref{integral eq0}). In the ordering adopted, this growth rate is small, $\gamma \ll \omega$, and the variational principle we have derived can be used to calculate the real part of the frequency. Knowing the latter, we now see what conclusions can be drawn about the growth rate using considerations of energy balance \cite{Proll}. Instead of doing this within the large-aspect-ratio approximation adopted in the previous section, we now consider general stellarator geometry, i.e., an arbitrary toroidal magnetic field with nested flux surfaces and finite global magnetic shear. 

Using the notation
		$$ \left\{ \cdots \right\} = \int_{-\infty}^\infty \frac{dl}{B} \int \left( \cdots \right) d^3v, $$
we first note that the work done by the electric field on the guiding centers of an arbitrary particle species $a$ is
	$$ 	- e_a \left\{ f_{a1} (v_\| {\bf b} + {\bf v}_{da} ) \cdot \nabla \phi \right\}
	= e_a  \left\{ \phi (v_\| {\bf b} + {\bf v}_{da} ) \cdot \nabla f_{a1} \right\}, $$
where ${\bf b} = {\bf B} / B$ and the adiabatic part of $f_{a1}$ does not contribute. If the fluctuating quantities are written
	$$ g_a({\bf R}) \sim {\rm Re} \; \hat g_a({\bf R}) e^{iS({\bf R})}, $$
	$$ \phi({\bf r}) \sim {\rm Re} \; \hat \phi({\bf r}) e^{iS({\bf r})}, $$
where ${\bf R}$ and $\bf r$ denote guiding-center and particle positions, respectively, the power transfer from  the fluctuating field to species $a$ is thus 
	$$ P_a = e_a {\rm Im} \left\{J_0 \hat \phi^\ast (i v_\| \nabla_\| \hat g_a - \omega_{da} \hat g_a) \right\}. $$
This work can be related to the potential energy
	$$ Q_a = e_a {\rm Im} \left\{J_0 \hat \phi^\ast \hat g_a \right\} $$
by noting that, according to the gyrokinetic equation (\ref{gk}), where we have dropped carets, 
\begin{align*}
& {\rm Im} \; \left\{ \left[ i v_\| \nabla g_a + (\omega - \omega_{da}) g_a \right] e_a J_0 \phi^\ast \right\} \\
	= & P_a + \omega_r Q_a  + \gamma {\rm Re} \left\{ e_a \phi^\ast J_0 g_a \right\} 
\end{align*}
is equal to
	$$ \gamma \frac{n_a e_a^2}{T_a} \int_{-\infty}^\infty \Gamma_0 (b) | \phi |^2 \frac{dl}{B}, $$
where $\omega = \omega_r + i \gamma$, and hence
\begin{align*}
 P_a = & - \omega_r Q_a - \gamma \bigg( {\rm Re} \left\{ e_a \phi^\ast J_0 g_a \right\} \\
	&- \frac{n_a e_a^2}{T_a} \int_{-\infty}^\infty \Gamma_0 (b) | \phi |^2 \frac{dl}{B} \bigg) 
\end{align*}
Summing over all species and using quasineutrality (\ref{qn}) gives an expression for the growth rate
	\bn \gamma \sum_a \frac{n_a e_a^2}{T_a} \int_{-\infty}^\infty [1 - \Gamma_0(b)] | \phi |^2 \frac{dl}{B} 
	= - \sum_a P_a, 
	\label{growth rate}
	\en
valid for all collisionless, electrostatic instabilities in arbitrary stellarator configurations. The right-hand side expresses how much power is transferred from each species to the turbulent fluctuations: any species with $P_a < 0$ is destabilizing, and vice versa. The quantity in Eq.~(\ref{growth rate}) is a ballooning-space version of the nonlinear electrostatic energy invariant of gyrokinetic theory \cite{Plunk}, and has sometimes been used in the past to estimate growth rates \cite{Rutherford}. 

If the conventional drift-wave ordering (\ref{drift-wave ordering}) is adopted, the energy transfer to the ions becomes
	$$ P_i = - \frac{e^2}{T_i} \left\{ |J_0 \phi|^2 \frac{\gamma}{(\omega_r - \omega_{di})^2 + \gamma^2}
	\omega_{di} (\omega_{*i}^T - \omega_{di}) f_{i0} \right\}, $$
where we have used Eq.~(\ref{gi}), 
and it is clear that the instability requires $\omega_{*i}^T \omega_{di} > 0$, at least in some parts of phase space. Near the marginal stability point, $\gamma \rightarrow 0+$, this expression reduces to 
	$$ P_i \rightarrow \frac{\pi e^2}{T_i} \left\{ |J_0 \phi|^2 \omega_{di} (\omega_{di} - \omega_{*i}^T)
	\delta (\omega - \omega_{di})f_{i0} \right\}, $$
and the electron contribution similarly becomes
	\bn P_e \rightarrow \frac{\pi e^2}{T_i}\left\{ |\overline{\phi}|^2 \overline \omega_{de} 
	(\overline \omega_{de} - \omega_{*e}^T)
	\delta (\omega - \overline \omega_{de})f_{e0} \right\}. 
	\label{Pe}
	\en
Both expressions can be understood as weighted averages of the quantity $\omega_{*a}^T \omega_{da}$, which needs to be positive, at least somewhere along the field line, in order for an instability to exist. 

More generally, whatever approximation is used when solving the gyrokinetic equation, if the right-hand side of Eq.~(\ref{growth rate}) turns out to be negative as $\gamma \rightarrow 0+$, there cannot exist any marginal stability point and therefore no instability. 
This argument can be made more precise \cite{Taylor-1968} by considering the Nyquist plot of
	$$ R(\omega) = \sum_a \int \frac{dl}{B}\left( \frac{n_a e_a^2 |\phi|^2 }{T_a} 
	- e_a \int \phi^\ast g_a J_0 \mathrm{d}^3v \right). $$
If 
	$${\rm Im} \; R(\omega) = - \sum_a Q_a $$
is negative for all {\em real} $\omega$, then the Nyquist contour cannot encircle the origin and there cannot be an instability. 

The utility of the variational principle derived in the previous section now becomes evident. If, for a given magnetic geometry, the frequency predicted by the principle, i.e., the solution $\omega$ to Eq.~(\ref{integral eq2}), has the opposite sign from the electron magnetic drift frequency $\overline{\omega}_{de}$, then there is no resonance in the denominator of the electron response (\ref{ge}) and no resonant power transfer from the electrons to the instability according to Eq.~(\ref{Pe}). The drift wave then propagates in the opposite direction from the electron precession, and there is no possibility of a collisionless TEM instability. The variational principle thus makes it possible to formulate geometry-dependent, sufficient criteria for TEM stability.

\section{Maximum-$J$ configurations}

To investigate the sign of $\omega_{*a} \omega_{da}$, we write the the magnetic drift frequency as
	$$ \omega_{da} = 
	\frac{v^2 }{\Omega_a} ({\bf k}_\perp \times {\bf b}) \cdot \left( \frac{1-\xi^2}{2} \nabla \ln B + \xi^2 \bfm{\kappa} \right), $$
where ${\bf b} = {\bf B} / B$, $\xi = v_\| / v$, and 
	$$ \bfm{\kappa} = {\bf b} \cdot \nabla {\bf b} = \nabla_\perp \ln B + \frac{\mu_0 p'(\psi)}{B^2} \nabla \psi $$
denotes the curvature of the magnetic field. Decomposing this vector and the wave vector as
	$$ \bfm{\kappa} = \kappa_\psi \nabla \psi + \kappa_\alpha \nabla \alpha, $$
	$$ {\bf k}_\perp = k_\psi \nabla \psi + k_\alpha \nabla \alpha, $$
and recalling the defintion of $\omega_{*a}$ gives
\begin{align*}
\omega_{*a} \omega_{da} = &\frac{v^2 n_a' T_a k_\alpha^2}{2 m_a n_a \Omega_a^2}
	\left[ (1 + \xi^2) \left( \kappa_\psi - \frac{k_\psi \kappa_\alpha}{k_\alpha} \right) \right.\\
	& \left.- \frac{\mu_0 p' (1-\xi^2)}{2B} \right]. 
\end{align*}
Specifically, for modes with $k_\psi = 0$, as is typical for interchanges, the product
	$$ \omega_{*a} \omega_{da} = \frac{v^2 n_a' T_a k_\alpha^2}{2 m_a n_a \Omega_a^2}
	\left[ (1 + \xi^2) \kappa_\psi  
	- \frac{\mu_0 p' (1-\xi^2)}{2B} \right] $$
consists of two terms, where the first term is destabilising if $\kappa_\psi < 0$, corresponding to ``bad'' curvature, and the second term is stabilizing for the usual orientation of the gradients, $n'_a < 0$ and $p' < 0$. By this token, the outboard side of a standard tokamak has unfavourable curvature, but the pressure gradient provides a stabilizing influence on electrostatic modes \cite{Rosenbluth-Sloan,Connor-1983}. 

We now turn to the bounce-averaged quantity $\omega_{*a} \overline{\omega}_{da}$, which according to Eq.~(\ref{Pe}) is the more relevant instability parameter for the electrons, or indeed any species with large thermal speed, $v_{Ta} \gg \omega / k_\|$. The bounce-average of the magnetic drift frequency is
	$$ \overline{\omega}_{da}= \overline{\textbf{v}_{da} \cdot( k_{\alpha} \nabla \alpha + k_\psi \nabla \psi ) }, $$
where \cite{Kadomtsev-trapped-particles,Rosenbluth-Sloan}
 	$$ \overline{\textbf{v}_{da} \cdot  \nabla \psi} = \frac{1}{e_a \tau_{ba}} \frac{\p J}{\p\alpha}, 
	$$
	$$ \overline{\textbf{v}_{da} \cdot  \nabla \alpha} = - \frac{1}{e_a \tau_{ba}} \frac{\p J}{\p\psi}, 
	$$
the bounce time is denoted by $\tau_{ba}\ = \tau_j / v$, and
	\bn {J}(E,\mu,\psi,\alpha) = \int mv_\| dl, 
	\label{J invariant}
	\en
is the parallel adiabatic invariant, with the integral taken between two consequtive bounce points. We regard $J$ as a function of the kinetic energy, $E = m_a v^2 / 2$ and magnetic moment $\mu = m_a v_\perp^2/2B$, as well as the field-line labels $\psi$ and $\alpha$. Well-optimized (omnigenous) stellarators have vanishing, or very small, bounce-averaged radial drift, $\overline{\textbf{v}_{da} \cdot  \nabla \psi} = 0$, implying that $J$ is constant on flux surfaces, $\p J / \p \alpha = 0$. In such configurations, we conclude that
	\bn \omega_{*a} \overline{\omega}_{da} 
	= -  \dfrac{k_{\alpha}^2T_a }{e_a^2  \tau_{ba}}\dfrac{d \ln n_a}{d \psi} \frac{\partial J}{\partial \psi} 
	\label{product}
	\en
is a negative quantity if $\p J / \p \psi < 0$ and $dn_a / d \psi < 0$. Such fields are called {\em maximum-J configurations} and were recognized already by Rosenbluth \cite{Rosenbluth} to have favourable stability properties (see also Ref.~\cite{Kadomtsev-trapped-particles}). 

Rosenbluth considered only axisymmetric systems, but for modern stellarator research the maximum-$J$ concept is of renewed importance because quasi-isodynamic \cite{HN,Nuhrenberg2010} stellarator designs tend to have this property, at least approximatively and particularly at high plasma $\beta$. This is in contrast to tokamaks, where the particles trapped on the outboard side of the torus generally have $\p J / \p \psi > 0$. 

Rosenbluth specifically considered isothermal plasmas with equal ion and electron temperatures, zero gyroradius and $\omega_{da} / \omega \ll 1$. In this case, he could demonstrate the absence of collisionless, electrostatic instabilities with low frequencies, $\omega \ll k_\| v_{Ti}$, so-called collisionless trapped-particle modes. From our analysis, we see that stability prevails well beyond this simple limit \cite{Proll}. If, for all species, $\omega \ll k_\| v_{Ta}$ and $0 < \eta_a < 2/3$, then $\omega^T_{*a} \overline{\omega}_{da} < 0$ everywhere in velocity space in a maximum-$J$ configuration, and the right-hand side of Eq.~(\ref{growth rate}) is negative for $\omega$ lying slightly above the real axis. All species are then stabilizing, $P_a > 0$, so there can be no marginal stability point and no instability, not only in Rosenbluth's limit but also for arbitrary values of $k_\perp \rho_a$, $\omega_{da} / \omega$, and finite temperature gradients up to $\eta_a = 2/3$ for all species. 

At higher frequencies, $\omega / k_\| \sim v_{Ti} \ll v_{Te}$, no such absolute statements can be made about stability, but it is possible to draw conclusions about the nature of any instability that could arise. If we fix our signs so that the ion diamagnetic frequency $\omega_{*i}$ is positive, then $\overline{\omega}_{de}$ will also be positive in a maximum-$J$ device (assuming that $dn_i/d\psi < 0$, as always), and $\omega_{*e}$ will be negative. If the electron temperature gradient is modest, $0 < \eta_e < 2/3$, then only modes propagating in the ion diamagnetic direction, $\omega > 0$, are able to interact resonantly with the precessing electrons, and these will be {\em stabilizing} according to Eq.~(\ref{Pe}), thus ruling out any instability that may reasonably be called a TEM. Moreover, if in addition the ion temperature gradient is not too large, $0 < \eta_i < 2/3$, one can show that the real frequency must be positive \cite{Proll}, so in this case there are absolutely no TEMs. Since these conclusions only rely on the maximum-$J$ condition and considerations of energy balance, they are independent of all other geometrical properties of the magnetic field. 

When the normalized temperature gradients exceed $\eta_a = 2/3$, it becomes more difficult to make general stability predictions, but the variational principle derived above makes it possible to draw conclusions that are more dependent on details of the magnetic geometry. First, if the aspect ratio is large and the number of trapped particles is small, then Eq.~(\ref{omega0}) predicts a frequency that is negative as long as $\eta_i < 1.64$, see Fig.~1. Therefore, in this limit, there is no resonance with the precessing electrons (in a maximum-$J$ device) and there should be no TEMs whatsoever, regardless of the electron temperature gradient. In the more realistic case of an ``order unity'' fraction of trapped particles, the real frequency is given by the variational form (\ref{variational form}) if the electron drift resonance is neglected in lowest order, and by Eq.~(\ref{integral eq2}) otherwise. In the former case, one sees that $\omega / \omega_{*e}$ is positive for small values of $b = (k_\perp \rho_i)^2$ and negative for large values of $b$, since the function 
	\bn F(b,\eta_i) = \Gamma_0(b) - \eta_i b [\Gamma_0(b) - \Gamma_1(b)] 
	\label{F}
	\en
decays (for some values of $\eta_i$ non-monotonically) with increasing $b$ as indicated by Eqs.~(\ref{low b}) and (\ref{high b}). Thus, the variational principle predicts that any drift-wave-type instability will propagate in the electron diamagnetic direction if the perpendicular wavelength is long and in the opposite direction if it is short. Only at short wavelengths is a resonant interaction with trapped electrons possible, and these will be stabilizing in a maximum-$J$ device.

\begin{figure}[htb]
\begin{center}
\includegraphics[width=0.45\textwidth]{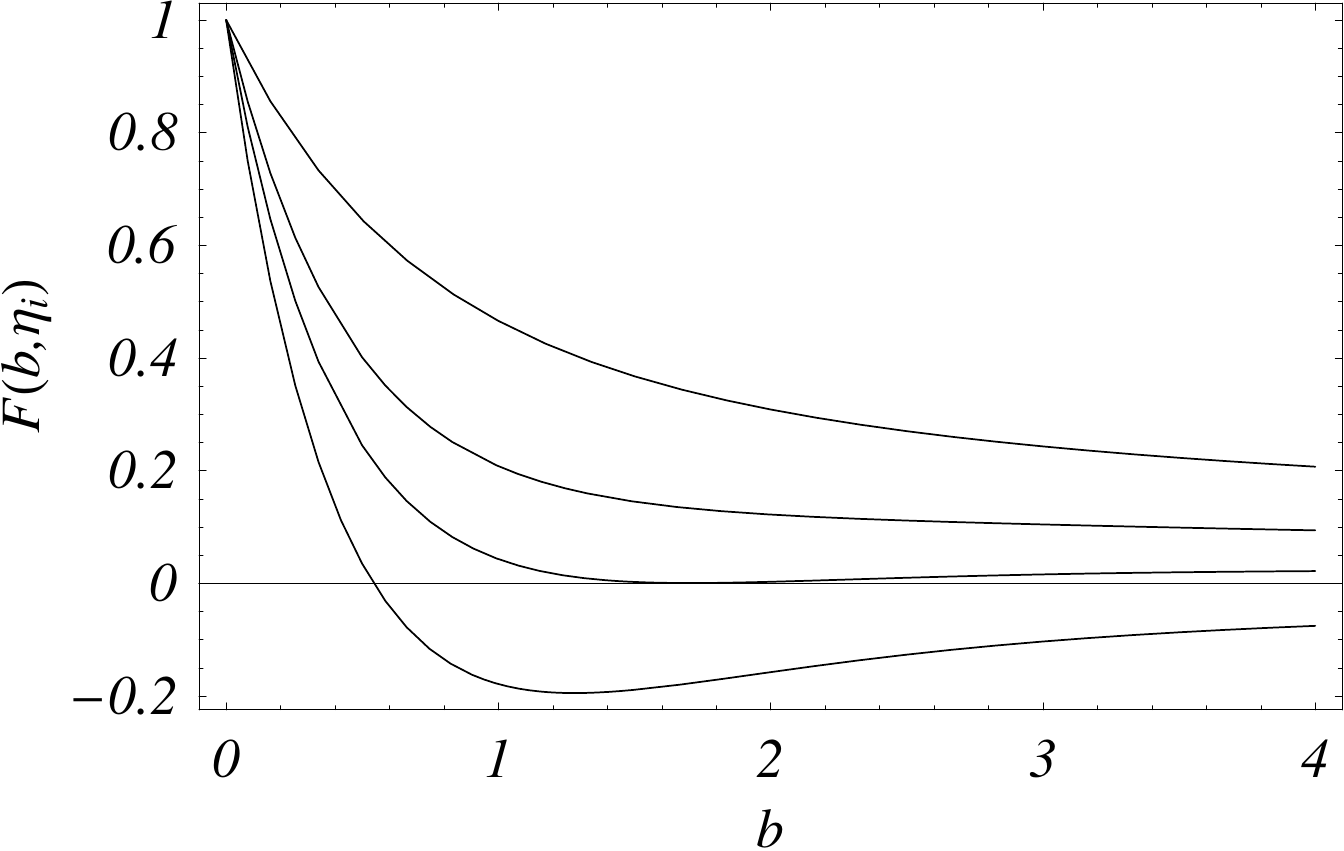}
\caption{\em The function $F(b,\eta_i)$ defined in Eq.~(\ref{F}) for various $\eta_i$. From above, the curves correspond to $\eta_i = 0$, 1, 1.64 and 2.5. }
\end{center}
\end{figure}

Such magnetic configurations are thus remarkably stable to TEMs, and a simple physical reason for this property was given in Ref.~\cite{Helander}. Consider any instability with $\omega \ll k_\| v_{Te}$ causing an electron to move the distance $\Delta \psi$ radially. Since $J$ is an adiabatic invariant, this movement must be accompanied by a change in energy, $\Delta E$, satisfying 
	$$ \Delta J = \frac{\p J}{\p \psi} \Delta \psi + \frac{\p J}{\p E} \Delta E = 0, $$
The electron thus gains the energy
	$$ \Delta E = - \frac{\p J / \p \psi}{\p J / \p E} \; \Delta \psi, $$
at the expense of the instability in question. Since $\p J/\p E > 0$, it follows that the condition $\p J / \p \psi < 0$ promotes stability if $dn/d\psi < 0$. It also follows that stability is not guaranteed for modes with frequencies high enough to be comparable to the electron bounce frequency. Since $\omega \sim \omega_{*e} \sim k_\perp \rho_i v_{Ti} / L_n$, where $L_n$ is the density length scale, we thus require
	$$ k_\perp \rho_i \ll k_\| L_n \sqrt{\frac{m_i}{m_e}} $$
for the stability properties to hold. The parallel wave number for TEMs can be estimated as $k_\| \sim N/R$, where $N$ is the number of periods and $R$ the major radius, so we need
	$$ k_\perp \rho_i \ll N\epsilon\sqrt{\frac{m_i}{m_e}},  $$
where the right-hand side is about 30 in Wendelstein 7-X. As will be seen in Part II of the present publication, the frequencies of the most unstable modes seen in gyrokinetic simulations are indeed much below the electron bounce frequency. 

All these analytical predictions are, strictly speaking, only valid in exactly omnigenous configurations. If the net radial drift does not vanish for some orbits, $\overline{\textbf{v}_{da} \cdot  \nabla \psi} \ne 0$, then the product
$\omega_{*a} \overline{\omega}_{da}$ in Eq.~(\ref{product}) acquires an additional term,
	$$ \omega_{*a} \overline{\omega}_{da} 
	= -  \dfrac{k_{\alpha}^2T_a }{e_a^2  \tau_{ba}}\dfrac{d \ln n_a}{d \psi} 
	\left( \frac{\partial J}{\partial \psi} - \frac{k_\psi}{k_\alpha} \frac{\p J}{\p \alpha} \right). $$
This term is of indefinite sign and can always be made negative by choosing $k_\psi$ appropriately. An instability with finite radial mode number, feeding off the density gradient of radially drifting trapped particls, is thus in principle possible. 

\section{Conclusions}

In summary, we conclude that collisionless trapped-particle instabilities can be very different in tokamaks and stellarators. In both types of devices, they reside in regions where trapped particles are present, and the instability drive is provided by ``bad'' curvature. In tokamaks, the regions of trapping and bad curvature usually overlap, but they need not do so in stellarators. The key instability parameter for the electrons, which is the product $\omega_{*e} \overline{\omega}_{de}$, is positive on the outboard side of a typical tokamak, signifying such an overlap. This implies that the trapped electrons precess in the same direction as electron drift waves, and therefore have the potential of destabilizing them, leading to collisionless TEMs. The opposite limit, where trapping is perfectly separated from bad curvature, is realized in maximum-$J$ configurations, where $\omega_{*e} \overline{\omega}_{de}$ is negative for all orbits. This condition is rarely satisfied exactly, but as we shall see in Part II, it can be true to a sufficiently good approximation in quasi-isodynamic stellarators, making TEMs much more stable than in a typical tokamak. 

In perfect maximum-$J$ configurations, it can be proved rigorously that the collisionless trapped-particle mode of Kadomtsev and Pogutse \cite{Kadomtsev-1967}, which is characterized by $\omega \ll k_\| v_{Ta}$ for all species, is stable for arbitrary wavelengths and density gradients as long as the temperature gradients satisfy $0 < \eta_a < 2/3$. Moreover, at higher frequencies, $\omega \sim k_\| v_{Ti}$, the electrons still exert a stabilizing influence  if $0 < \eta_e < 2/3$, so there can be no unstable collisionless TEMs. There may be other instabilities than TEMs present in the plasma, but they must be drawing energy from the ions rather than the electrons. Furthermore, as we have seen from a variational principle for the real mode frequency, the  stability window for TEMs is, in several situations, even larger than these results suggest. In particular, it appears that the destabilizing potential of the electron temperature gradient is very limited. 

Since much of the analytical theory presented here strictly only applies in the idealized limit of perfect maximum-$J$ geometry, it is pertinent to ask how well the predictions are borne out in practice, in real stellarators. This question can only be answered by numerical simulations and is the topic of Part II. 

\begin{acknowledgments}
The authors gratefully acknowledge very helpful discussions with Jack Connor. 
\end{acknowledgments}

\begin{references}
\bibitem{Kadomtsev-1967} B.B. Kadomtsev and O.P. Pogutse, Sov. Phys. JETP {\bf 24},
1172 (1967).

\bibitem{Rosenbluth} M.N. Rosenbluth, Phys. Fluids {\bf 11}, 869 (1968).

\bibitem{Kadomtsev-DTEM} B.B. Kadomtsev and O.P. Pogutse, Sov. Phys. Dokl. {\bf 14}, 470 (1969). 

\bibitem{Kadomtsev-trapped-particles} B.B. Kadomtsev and O.P. Pogutse, Nucl. Fusion {\bf 11}, 67 (1971).

\bibitem{Adam} J.C. Adam, W.M. Tang and P.H. Rutherford, Phys. Fluids {\bf 19}, 561 (1976).

\bibitem{Ross} D.W. Ross, W.M. Tang and J.C. Adam, Phys. Fluids {\bf 20}, 613 (1977).

\bibitem{Catto} P.J. Catto and K.T. Tsang, Phys. Fluids {\bf 21}, 1381 (1978). 

\bibitem{Rewoldt} G. Rewoldt, W.M. Tang and E.A. Frieman, Phys. Fluids {\bf 24}, 238 (1981).

\bibitem{Berk} H. L. Berk, M. N. Rosenbluth, R. H. Cohen, and W. M. Nevins, Phys. Fluids {\bf 28}, 2824 (1985). 

\bibitem{Kendl} A. Kendl, Plasma Phys. Control. Fusion {\bf 43} 1559 (2001). 

\bibitem{Pax-2007} P. Xanthopoulos and F. Jenko, Phys. Plasmas {\bf 14}, 042501 (2007).

\bibitem{Baumgaertel}  J. A. Baumgaertel, G. W. Hammett, D. R. Mikkelsen, M. Nunami, and P. Xanthopoulos, Phys. Plasmas {\bf  19}, 122306 (2012).

\bibitem{Proll} J.H.E. Proll, P. Helander, J.W. Connor and G.G. Plunk, Phys. Rev. Lett. {\bf 108}, 245002 (2012). 

\bibitem{Helander} P. Helander, C.D. Beidler, T.M. Bird, M. Drevlak, Y. Feng, R. Hatzky, F. Jenko, R. Kleiber, J.H.E. Proll, Yu. Turkin and P. Xanthopoulos, Plasma, Phys. Control. Fusion {\bf 54}, 124009 (2012).

\bibitem{Helander-2007} P. Helander, Phys. Plasmas {\bf 14}, 104501 (2007). 

\bibitem{Sugama-2011} H. Sugama, T.-H. Watanabe, M. Nunami, and S. Nishimura, Phys. Plasmas {\bf 18}, 082505 (2011). 

\bibitem{Catto} P.J. Catto, Plasma Phys. {\bf 20}, 719 (1978).

\bibitem{Dewar-Glasser} R.L. Dewar and A.H. Glasser, Phys. Fluids {\bf 26}, 3038 (1983). 

\bibitem{Helander-Sigmar} P. Helander and D.J. Sigmar, {\em
Collisional transport in magnetized plasmas} (Cambridge University
Press, 2002).

\bibitem{Antonsen-Lane} T.M. Antonsen and B. Lane, Phys. Fluids {\bf 23}, 1205 (1980).

\bibitem{Catto-Tang-Baldwin} P.J. Catto, W.M. Tang and D.E. Baldwin, Plasma Phys. {\bf 23}, 639 (1981).

 \bibitem{Connor-TEM} J.W. Connor, R.J. Hastie and P. Helander, Plasma Phys. Control. Fusion {\bf 48}, 885 (2006).

\bibitem{Plunk} G.G. Plunk, T. Tatsuno and W. Dorland, New J. Phys. {\bf 14}, 103030 (2012). 

\bibitem{Rutherford} P.H. Rutherford and E.A. Frieman, Phys. Fluids {\bf 11}, 569 (1968).

\bibitem{Taylor-1968} J.B. Taylor and R.J. Hastie, Plasma Phys. {\bf 10}, 479 (1968). 

\bibitem{Rosenbluth-Sloan} M.N. Rosenbluth and M.L. Sloan, Phys. Fluids {\bf 14}, 1725 (1971). 

\bibitem{Connor-1983}  J.W. Connor, R.J. Hastie and T.J. Martin, Nucl. Fusion {\bf
23}, 1702 (1983).

\bibitem{HN} P. Helander and J. N\"uhrenberg, Plasma Phys. Control. Fusion {\bf 51},
055004 (2009). 

\bibitem{Nuhrenberg2010} J. N\"uhrenberg, Plasma Phys. Control. Fusion {\bf 52},
124003 (2010).

\bibitem{Taylor-1964} J.B. Taylor, Phys. Fluids {\bf 7}, 767 (1964).

\bibitem{Rutherford-Frieman} P.H. Rutherford and E.A. Frieman, Phys. Fluids {\bf 11}, 252 (1968). 
\end{references}
\section*{Appendix}

Over the years, the stability criterion $\p J / \p \psi < 0$ has surfaced in various guises in the literature. 
In this Appendix, we establish the relation to a classic stability criterion derived by Taylor, Rutherford and Frieman in the 1960's. Taylor \cite{Taylor-1964} considered flute modes in mirror machines in the zero-gyroradius-limit, where he wrote the equilibrium distribution function in $(\psi,\alpha,J,\mu)$ space as $F[\mu,J,E(\psi,\alpha,\mu,J)]$, where $E$ is the particle energy, and estabilished that a sufficient criterion for stability against electrostatic modes is
	\bn \left( \frac{\p F}{\p E} \right)_{\mu,J}< 0. 
	\label{JBT criterion}
	\en
Rutherford and Frieman derived a similar criterion for configurations where all field lines close on themselves, again in the drift kinetic limit \cite{Rutherford-Frieman}. Since the phase-space volume element (integrated over the gyroangle) is
	$$ \mathrm{d}^3r \mathrm{d}^3v = \frac{4 \pi}{m^2 | v_\| |} \mathrm{d}E \mathrm{d}\mu \mathrm{d}\psi \mathrm{d}\alpha \mathrm{d}l, $$
and 
	$$ \left( \frac{\p J}{\p E} \right)_\mu = \frac{\p}{\p E} \int \sqrt{2 m (E - \mu B - e \phi)} \; \mathrm{d}l = \int \frac{\mathrm{d}l}{v_\|}, $$
the integral of $F$ is
	$$ \int F \mathrm{d}\mu \mathrm{d}J \mathrm{d}\psi \mathrm{d}\alpha = \frac{m^2}{4 \pi} \int F \; \mathrm{d}^3r \mathrm{d}^3v. $$
and it is clear that $F$ is simply proportional to our distribution function $f$. Furthermore, if we regard $J$ as a function of $(E,\mu,\psi)$ in an omnigeneous field (where $\p J / \p \alpha = 0$) and write $F = F[E,\mu,\psi(E,\mu,J)]$, then
\begin{align*}
\left( \frac{\p F}{\p E} \right)_{\mu,J} & = 
	\left( \frac{\p F}{\p E} \right)_{\mu, \psi} + \left( \frac{\p F}{\p \psi} \right)_{E,\mu} \left( \frac{\p \psi}{\p E} \right)_{\mu,J}\\
	& = - \frac{F}{T} - \frac{\p J / \p E}{\p J / \p \psi} \left( \frac{\p F}{\p \psi} \right)_{E,\mu}
\end{align*}
if $F$ is Maxwellian with temperature $T$. Since $\p J / \ \p E > 0$, the criterion (\ref{JBT criterion}) is satisfied if $\p J / \p \psi < 0$ and
	$$ \frac{\p F}{\p \psi} = F \left[ \frac{\mathrm{d} \ln n}{\mathrm{d} \psi} + \left( \frac{E}{T} - \frac{3}{2} \right) \frac{\mathrm{d} \ln T}{\mathrm{d} \psi} \right] < 0. $$
This is the case for all $E$ if $0 < \mathrm{d}\ln T / \mathrm{d} \ln n < 2/3$, as found in Sec.~VI above. 

\end{document}